\documentclass[prl,showpacs,superscriptaddress,preprintnumbers,amsmath]{revtex4}
\usepackage[dvips,final]{graphicx}
  \usepackage{amssymb}
   \usepackage{amsmath}
    \usepackage{epsfig}
     \usepackage{bm}% bold math
      \usepackage{pifont}
\newcommand{\vecc}[1]{\mbox{\boldmath $#1$}}

\def\ve{\varepsilon}

\def\ex{\hbox{e}}
\def\<{\langle}
\def\>{\rangle}

\def\({\left(}
\def\[{\left[}
\def\){\right)}
\def\]{\right]}

\begin{document}
\preprint{\hbox{RUB-TPII-14/09}}
%\vspace*{-10mm}
%%%%%%%%%%%%%%%%%%%%%%%%%%%%%%%%%%%%%%%%%%%%%%%%%%%%%%%%%%%%%%%%%%%%%%%%

\title{Understanding the evolution of transverse-momentum
       dependent parton densities}
\author{I.~O.~Cherednikov\footnote{Also at: {\sl ITPM,
             Moscow State U., RU-119899, Moscow, Russia}} }
\email{igor.cherednikov@jinr.ru}
\affiliation{INFN Cosenza,
             Universit$\grave{a}$
             della Calabria,  I-87036 Rende (CS), Italy}
\affiliation{Bogoliubov Laboratory of Theoretical Physics,
             JINR,
             RU-141980 Dubna, Russia}
%\affiliation{Institute for Theoretical Problems of Microphysics,
%             Moscow State University, \\ RU-119899, Moscow, Russia}
\author{N.~G.~Stefanis}
\email{stefanis@tp2.ruhr-uni-bochum.de}
\affiliation{Institut f\"{u}r Theoretische Physik II, \\
             Ruhr-Universit\"{a}t Bochum,
             D-44780 Bochum, Germany}
\affiliation{Bogoliubov Laboratory of Theoretical Physics,
             JINR,
             RU-141980 Dubna, Russia}

\date{\today}

\begin{abstract}
Different approaches to define transverse-momentum dependent parton
distribution functions are considered from the point of view of their
renormalization-group properties.
The associated one-loop anomalous dimensions of these quantities are
presented and compared to each other.
We give arguments in favor of the ``pure light-like'' definition,
and the use of the light-cone gauge.

\end{abstract}


\maketitle

%%%%%%%%%%%%%%%%%%%%%%%%%%%%%%%%%%%%%%%%%%%%%%%%%%%%%%%%%%%%%%%%%%%%%%%

Transverse-momentum dependent parton densities, or distribution
functions (TMD PDFs), are used in the study of semi-inclusive
hadronic reactions, e.g., semi-inclusive deep-inelastic scattering
(SIDIS), or the Drell-Yan (DY) process,
$
     \(\gamma^*(q) + { h_1(P)} \to {h_2 (P')} + {\cal X} \ , \
     h_1 (P) + { h_2(P')}  \to   {\cal X} \),
$
in order to describe the inner structure of hadrons by taking into
account the longitudinal, as well as the transversal, partonic degrees
of freedom \cite{Sop77, CS81, Col03, BR05}.
The explicit {\it operator definition} of TMD PDFs is quite
problematic, in particular, due to a number of extra (in contrast to the
ordinary {\it integrated} distributions) singularities, and is nowadays
under active investigation (see, e.g.,
\cite{Col03, JY02, BJY03, BMP03, CT05, CRS07, CS07, Bacch08, Col08}).
In the present work, we discuss different operator definitions of TMD
PDFs from the point of view of their {\it renormalization-group} (RG)
properties.

Let us start with three different definitions of a {\it general}
unintegrated quark distribution:
\begin{eqnarray}
  && \tilde {\cal F}_{[n]} \(x, \vecc k_\perp; \mu, \eta \)
= \frac{1}{2}
  \int \frac{d\xi^- d^2\xi_\perp}{2\pi (2\pi)^2}\
  \ex^{-ik^{+}\xi^{-} +i \mbox{\footnotesize\boldmath$k_\perp$}
\cdot
\mbox{\footnotesize\boldmath$\xi_\perp$}}
  \< \,
              h \, | \bar \psi (\xi^-, \vecc \xi_\perp) \
              [\xi^-, \vecc \xi_\perp;
   \infty^-, \vecc \xi_\perp]_n^\dagger
   \nonumber \\
  &&
   [\infty^-, \vecc\xi_\perp;
   \infty^-, \vecc \infty_\perp]_{\vecc l}^\dagger
   [\infty^-, \vecc \infty_\perp;
   \infty^-, \vecc 0_\perp]_{\vecc l}
   [\infty^-, \vecc 0_\perp; 0^-,\vecc 0_\perp]_n \
   \psi (0^-,\vecc 0_\perp) \, | \, h
   \> \ \, ,
\label{eq:tmd_n}
\end{eqnarray}
%Eq (1)

\begin{eqnarray}
  && \tilde {\cal F}_{[v]} \left(x, \vecc k_\perp;\mu, \zeta \right) =
  \frac{1}{2}
  \int \frac{d\xi^- d^2\xi_\perp}{2\pi (2\pi)^2}\
  \ex^{-ik^{+}\xi^{-} +i \mbox{\footnotesize\boldmath$k_\perp$}
\cdot \mbox{\footnotesize\boldmath$\xi_\perp$}}
  \< \,
              h \, | \bar \psi (\xi^-, \vecc \xi_\perp) \
              [\xi^-, \vecc \xi_\perp;
   \infty^-, \vecc \xi_\perp]_v^\dagger
 \nonumber \\
&& \times
   [\infty^-, \vecc\xi_\perp;
   \infty^-, \vecc \infty_\perp]_{\vecc l}^\dagger
   [\infty^-, \vecc \infty_\perp;
   \infty^-, \vecc 0_\perp]_{\vecc l}
   [\infty^-, \vecc 0_\perp; 0^-,\vecc 0_\perp]_v \
   \psi (0^-,\vecc 0_\perp) \, | \, h
   \> \ \, ,
\label{eq:tmd_v}
\end{eqnarray}
%Eq (2)

\begin{eqnarray}
  && \tilde {\cal F}_{[v_0]} \left(x, \vecc k_\perp;\mu, \zeta_0 \right)
  =
\frac{1}{2}
  \int \frac{d\xi^- d^2\xi_\perp}{2\pi (2\pi)^2}\
  \ex^{-ik^{+}\xi^{-} +i \mbox{\footnotesize\boldmath$k_\perp$}
\cdot \mbox{\footnotesize\boldmath$\xi_\perp$}}
  \< \,
              h \, | \bar \psi (\xi^-, \vecc \xi_\perp) \
   [\xi^-, \vecc \xi_\perp;
   0^-, \vecc 0_\perp]_{v_0} \
   \psi (0^-,\vecc 0_\perp) \, | \, h
   \> \ \, ,
\label{eq:tmd_v0}
\end{eqnarray}
%Eq (3)
where in all cases gauge invariance is ensured by means of the
path-ordered contour-dependent Wilson-line operators (gauge links)
with the generic form
\begin{equation}
   \[y,x \]_{r}
=
  {\cal P} \exp
  \left[-ig\int_{\tau_1}^{\tau_2}d\tau r^{\mu} A^{a}_{\mu}(r \tau) t^{a}
  \right] \ \  , \  \ r^\mu \tau_1 = x \ , \  r^\mu \tau_2 = y \ ,
\label{eq:link}
\end{equation}
%Eq (4)
and one has to distinguish between longitudinal
$[ \ , \ ]_{[n,\, v, \, v_0]}$ and
transversal $[ \ , \ ]_{[\, \vecc l\, ]}$
gauge links \cite{BJY03, BMP03}.
In general, the state $| h \>$ is a hadron with momentum $P$ and
spin $S$, but for the sake of simplicity we restrict ourselves in what
follows to the ``distribution of a quark in a quark with momentum
$p$''---sufficient for the investigation of the ultraviolet (UV) behavior
(omitting color and flavor indices).
The transverse gauge links, extending to light-cone infinity, are key
elements in Eqs.\ (\ref{eq:tmd_n})--(\ref{eq:tmd_v0}) in order to
ensure full gauge invariance, while the dependence on the hard momentum
scale(s) (e.g.,  $Q^2$) is taken into account via appropriate
evolution equation(s) (more later).
Below, we present the results for the functions
(\ref{eq:tmd_n})--(\ref{eq:tmd_v0}), when projected onto the
$\gamma^+$-matrix:
$
  {\cal F}
  \equiv
  \frac{1}{2}\ {\rm Tr} \[\gamma^+ \tilde {\cal F} \] \ .
$
Additional variables are introduced in order to regularize the extra
singularities of the TMD PDFs, viz.,
\begin{equation}
\zeta = \frac{4(P\cdot v)^2}{v^2} \ , \ \ \ \
\zeta_0 = \frac{4(P\cdot v_0)^2}{v_0^2} \ ,
\end{equation}
%Eq (5)
whereas $\eta$ is a light-cone regularization parameter to be defined
later.
Note that definition (\ref{eq:tmd_v0}) involves the ``direct''
gauge link between two points separated by an essentially
non-lightlike distance.
From the point of view of kinematics and power-counting,
this is forbidden and seems to have nothing to do with the
(factorized) semi-inclusive processes under consideration.
However, we have included it in our list because it can be used in the
analysis of certain model approaches, or in lattice simulations
(see the recent papers \cite{Hae09, Hae09_1, Musch09}).

The definition given by Eq.\ (\ref{eq:tmd_n}) contains gauge links
pointing in the light-cone directions, and depending on the hard
scale $Q^2$.
This definition is consistent, at least naively, with the corresponding
collinear factorization (though a formal proof is lacking).
However, this definition, taken literally, produces---beyond the
tree-level---divergences, as actually expected.
In total, the singularities arising (in the one-loop order) in the
definition (\ref{eq:tmd_n}) belong to one of the following three
classes:

\begin{enumerate}

  \item Usual {\it UV-singularities} $\sim \frac{1}{\ve}$ from loop
  integrations, which can be removed by using the standard
  $R-$operation and are controlled by renormalization-group evolution
  equations, e.g., in the fully integrated case it amounts to the
  DGLAP equation.

  \item Pure {\it rapidity divergences}, which appear only in the
  {\it unintegrated} case.
  They cancel in the integrated distributions, but they are present in
  the TMD case giving rise to logarithmic and double-logarithmic terms
  of the form $\sim \ln \zeta \ , \ \ln^2 \zeta$; they have to be
  resumed by a consistent procedure.
  \item {\it Overlapping divergences}, which contain both UV and
  soft singularities simultaneously.
  They are highly undesirable, since they break the correct
  UV-evolution, and furthermore depend on the parameters of the chosen
  gauge, thus invalidating the definition of the TMD PDFs with respect
  to full gauge invariance.
  In this case, a generic singular term has the form
  $
  \sim \frac{1}{\ve} \ \ln \zeta \ ,
  $
  meaning that the UV pole $\ve^{-1}$ mixes with a ``soft''
  divergence, regularized by the auxiliary parameter $\zeta$.
  This prevents the removal of {\it all} UV-singularities by the
  standard $R-$procedure, making it necessary to apply a special
  {\it generalized} renormalization procedure.

\end{enumerate}
The latter two classes originate, in fact, from the uncompensated
light-cone artifacts, which stem either from the lightlike gauge links
(in covariant gauges), or from specific terms in the gluon propagator
in the (singular) light-cone axial gauges.
Let us emphasize that, while the singularities of the third class may
be simply regularized by some (rapidity) cutoff, which is ``separated''
from other variables, the effect of the second class is more severe,
because these singularities affect the UV-renormalization procedure,
change the anomalous dimensions, and modify, therefore, the RG-evolution.

In order to avoid the above-mentioned problems, the following approaches
have been proposed in the literature:

\begin{enumerate}

\item Shift in covariant gauges the gauge links off the light-cone:
$v^2 < 0 \ , \ v^+ \ll v^- $,  or use instead the non-lightlike axial
gauge $(v \cdot A) = 0 \ , \ v^2 < 0$ \cite{CS81}.
This amounts to definition (\ref{eq:tmd_v}).

\item Stay on the light-cone, Eq. (\ref{eq:tmd_n}), but subtract some
specific soft factor $R$, which is defined in such a way as to exactly
cancel the extra divergences \cite{CH00, Hau07, CM04}.
Thus, definition (\ref{eq:tmd_n}) is substituted by the ``subtracted''
function ${\cal F}_{[n]} \to {\cal F}_{[n]} \cdot R^{-1}$.

\item Perform a direct regularization of the light-cone singularities
in the gluon propagator \cite{CFP80}
\begin{equation}
\frac{1}{q^+} \to \frac{1}{[q^+](\eta)} \ ,
\label{eq:reg_eta}
\end{equation}
%Eq(5-1)
where $\eta$ is an additional dimensional parameter \cite{CS07}.
In this case, a generalized renormalization is in order, which is
formally equivalent to multiplying the TMD PDF by a particular soft
factor \cite{KR87}:
\hbox{${\cal F}_{[n]} (\eta) \to {\cal F}_{[n]} (\eta) \cdot R^{-1} (\eta)$}.
The introduction of the small parameter $\eta$ allows one to keep the
overlapping singularities under control and treat the extra term in the
UV-divergent part by means of the {\it cusp anomalous dimension}, which
in turn determines the specific form of the gauge contour in the soft
factor $R$.

\item Still use the light-cone axial gauge, but supply it with the
Mandelstam-Leibbrandt pole prescription
\cite{Man83, Lei84, LN83, BDS87, BKKN93, BA96, BHKV98}:
\begin{equation}
\frac{1}{q^+} \ \to \  \frac{1}{q^+ + i0q^-}  \ \ \hbox{or} \ \  \frac{q^-}{q^+q^- + i0}
  \ .
\label{eq:reg_ml}
\end{equation}
%Eq(5-2)
Now the overlapping singularities do not appear at all, at least at the
level of the one-loop order, while the contribution of the soft factor
is reduced to unity, rendering the gauge-invariant definition valid
\cite{CS09}.

\end{enumerate}

Let us now list the UV-renormalization-group equations for the above
definitions.
The off-the-light-cone TMD PDFs (\ref{eq:tmd_v}) and (\ref{eq:tmd_v0})
do not contain overlapping singularities.
Therefore, the only source to produce their UV-divergences, when the
two quark field operators are separated by a non-lightlike distance,
are the divergences of these operators themselves and those entailed by
the non-lightlike gauge links.
Therefore, the renormalization-group equation reads
\cite{KMPLA89, JMY04}
\begin{equation}
  \mu \frac{d }{d\mu} \ {\cal F}_{[v\, , \, v_0]}
  =
   \gamma_{\rm LC} \  {\cal F}_{[v\, , \, v_0]}\  \ , \ \
   \gamma_{\rm LC} = \frac{3}{4} \ \frac{\alpha_s C_{\rm F}}{\pi}
   + O(\alpha_s^2) \ ,
\label{eq:uv_v}
\end{equation}
%Eq (6)
where $\gamma_{\rm LC}$ is the anomalous dimension of the two-fermion
operator in the light-cone gauge.
If one factorizes out the soft contribution $R_v$, as it was proposed in
Ref. \cite{JMY04}, then the anomalous dimension changes and one has
\begin{equation}
 {\cal F}_{[\, v \,]} \to {\cal F}_{[\, v \,]} \cdot R_v^{-1} \ \ , \ \
  \mu \frac{d }{d\mu} \ \[ {\cal F}_{[\, v\,]}\cdot R_v^{-1} \]
  =
   ( \gamma_{\rm LC} - \gamma_{\rm R} )\ \[ {\cal F}_{[\, v\,]}
   \cdot R_v^{-1} \]\ ,
\label{eq:uv_vr}
\end{equation}
%Eq (7)
where $\gamma_{\rm R}$ is the one-loop anomalous dimension of the soft
factor $R_v$.

In contrast, the anomalous dimension of the ``light-cone'' TMD PDF, before
subtraction, deviates from $\gamma_{\rm LC}$ and this deviation is
determined by the cusp anomalous dimension \cite{CS07}:
\begin{equation}
  \mu \frac{d }{d\mu} \ {\cal F}_{[\, n\,]}
  =
   ( \gamma_{\rm LC} - \gamma_{\rm cusp}  ) \  {\cal F}_{[\, n\,]} \ .
\label{eq:uv_n}
\end{equation}
%Eq (8)
Hence, the generalized renormalization procedure restores the ``broken''
anomalous dimensions, so that one finds
\begin{equation}
  {\cal F}_{[n]} (\eta) \to {\cal F}_{[n]} (\eta) \cdot R^{-1} (\eta) \ \ , \ \
  \mu \frac{d }{d\mu} \ \[ {\cal F}_{[\, n\,]} \cdot R_n^{-1} \]
  =
    \gamma_{\rm LC} \  \[ {\cal F}_{[\, n\,]} \cdot R_n^{-1}  \]\ .
\label{eq:uv_nr}
\end{equation}
%Eq (9)

In the light-cone gauge with the Mandelstam-Leibbrandt prescription,
definition (\ref{eq:tmd_n}) yields an anomalous dimension without
lightlike artifacts from the very beginning, i.e.,
\begin{equation}
  \mu \frac{d }{d\mu} \ \[ {\cal F}_{[\, n\,]}^{\rm ML} \cdot R_n^{-1} \]
  =
  \mu \frac{d }{d\mu} \  {\cal F}_{[\, n\,]}^{\rm ML}
  =
    \gamma_{\rm LC} \  \[ {\cal F}_{[\, n\,]}^{\rm ML} \cdot R_n^{-1} \]
  =
    \gamma_{\rm LC} \ {\cal F}_{[\, n\,]}^{\rm ML}\ .
\label{eq:uv_ml}
\end{equation}
%Eq (10)

\paragraph{Conclusions.} We presented the one-loop UV anomalous dimension of the
TMD PDFs, defined in different ways expressed via
Eqs.\ (\ref{eq:tmd_n}-\ref{eq:tmd_v0}).
These anomalous dimensions can be used to construct corresponding
evolution equations, while the evolution in the {\it rapidity}
variable---either $\zeta$, or $\eta$, depending on the approach
applied---constitutes a separate task which will be considered
elsewhere.
We have also shown that using the ``completely lightlike'' definition
(\ref{eq:tmd_n}) in the light-cone gauge in conjunction with the
Mandelstam-Leibbrandt pole prescription appears to be the most
economic approach in the sense that it avoids---at least in the
one-loop order---undesirable overlapping divergences.

\paragraph{Open questions.} Let us sketch a couple of important and
still unsolved problems.

$(i)$ A major task within the TMD approach concerns the
{\it factorization} of semi-inclusive processes.
An all-order factorization (in a covariant gauge) was studied in Ref.
\cite{JMY04}, but in this work, the definition (\ref{eq:tmd_v}) was
used which contains off-the-light-cone gauge links.
No explicit proof of a factorization theorem with pure lightlike TMD
PDFs is known at present.

$(ii)$ Another problem pertains to the relationship between
unintegrated TMD PDFs and Feynman distribution functions, the latter
appearing in factorized inclusive processes (e.g., DIS).
After integrating over the transverse momentum $\vecc k_\perp$ of a
parton, one would expect that the standard integrated distribution is
reproduced from the TMD PDF.
It has been shown in Refs.\ \cite{CS07, CS09} that the
``pure light-like'' definition (\ref{eq:tmd_n}) does indeed yield,
after integration, an $x-$dependent distribution function that obeys
the DGLAP evolution equation:
\begin{equation}
  \int\! d^2 \vecc k_\perp \ {\cal F}_{[\, n\, ]} (x, \vecc k_\perp, \mu)
  =
  F_{[\, n\, ]} (x, \mu) \ , \  \mu \frac{d}{d\mu} \ F_{[\, n\, ]}
  =
  {\cal K}_{\rm DGLAP} \otimes F_{[\, n\, ]} \ .
\end{equation}
%Eq (13)
The reason is that the regularization via (\ref{eq:reg_eta}) and
(\ref{eq:reg_ml}) does not break the light-cone properties of the
function and, whence, the overlapping singularities in the real and
virtual gluon contributions cancel against each other after the
$\vecc k_\perp$-integration.
On the other hand, performing the $\vecc k_\perp$-integral in the
function (\ref{eq:tmd_v}), one cannot even expect to obtain a
distribution having DGLAP evolution.
In contrast, one gets a function containing off-the-light-cone gauge
links along the vector $v = (v^+, v^-, \vecc 0_\perp)$, i.e.,
\begin{equation}
  \int\! d^2 \vecc k_\perp \ {\cal F}_{[\, v\, ]} (x, \vecc k_\perp, \mu)
  =
  F_{[\, v\, ]} (x, \mu) \ \ , \  \ \mu \frac{d}{d\mu} \ F_{[\, v\, ]}
  =
  {\cal K}_{v} \otimes F_{[\, v\, ]} \ \ , \ \
  {\cal K}_{v} \neq {\cal K}_{\rm DGLAP} \ .
\end{equation}
%Eq (14)
The RG-properties of this object differ from those of the distribution
with lightlike gauge links (which fulfills the DGLAP equation), the
reason being that the latter produce specific UV singularities without
a possibility to perform a regular transition from off-the-light-cone
gauge links back to pure light-cone gauge links (see, e.g., Ref.
\cite{BKKN93}).
Thus, applying definition (\ref{eq:tmd_v}), one should keep in mind
that the relation to the integrated distribution is, at least, obscure.

\acknowledgments
This work was supported in part by the Heisenberg-Landau Program 2009
and the Russian Federation Scientific Schools grant 195.2008.9.

\end{document}